\documentclass[aps,prl,twocolumn,showpacs,groupedaddress]{revtex4}

\usepackage{graphicx}

\begin{document}

\title{Landau-Zener transition of a two-level system driven by  spin chains near their critical points}
\author{L. C. Wang, X. L. Huang, and X. X. Yi} \email{yixx@dlut.edu.cn}
\affiliation{School of Physics and Optoelectronic Technology,
              Dalian University of Technology, Dalian 116024, China}

\date{\today}

\begin{abstract}
The Landau-Zener(LZ) transition of a two-level system
coupling to spin chains near their critical points is studied in
this paper. Two kinds of spin chains, the Ising spin chain and XY
spin chain, are considered. We calculate and  analyze the effects of
system-chain coupling on the LZ transition.  A relation between the
LZ transition and the critical points of the spin chain is
established. These results suggest that LZ transitions may serve as
the witnesses of criticality of the spin chain. This may provide a
new way to study quantum phase transitions as well as LZ transitions.
\end{abstract}

\maketitle

\section{Introduction}
Quantum information processing promotes renewed attentions in
quantum two-level systems in recent years. A number of two-level
systems have been tested as good candidates of qubits that are the
least units in quantum information processing \cite{neilsen}. A good
qubit requires that the qubit is well isolated from its environments
and easy to manipulate. There are many ways to manipulate qubits,
one of them is to use Landau-Zener (LZ) sweeps, which have been
realized in recent experiments on superconducting qubits
\cite{izmalkov2004, oliver2005, sillanpaa2006}.

The LZ transition  occurs when two instantaneous eigenvalues of a
quantum  system come close together due to the parameter change. It
has attracted attentions for decades since the work in the early
1930s in slow atomic collisions \cite{landau1932, zener1932,
stueckelberg1932} and spin dynamics \cite{majorana1932}. The LZ
theory has found many applications, such as the mentioned
manipulation  of qubits, the enhancement of  the macroscopic quantum
tunneling \cite{ankerhold2003} and the  control of the transition
probability as well as  the quantum phase factor \cite{saito2004}.
Recently, the LZ effect  was proved to be useful in quantum
information processing such as the preparation of  quantum states in
circuit QED systems \cite{saito2006},  creating  entangled modes in
cavities \cite{wubs2007} and fault-tolerant single-qubit gate
operations \cite{hicke2006}. In practice,  the qubit is always
influenced by its environment, leading to decoherence in that
system. Therefore, taking the environment into account when study
the LZ transition is a practical consideration. The LZ transitions
in two-state systems dissipatively coupled to their environments
have been studied extensively \cite{kayanuma1987, wubs2006,
saito2007,wubs2005,wan2007}, those results  show that the two-level
system is robust against the influence of environment.  Those
studies do not consider the  interactions among the particles in the
environment, hence the information of the particle-particle
correlation can not reflected in the LZ transitions. It is well
known that the spin-spin coupling in the spin-chain may result in
quantum phase transitions, due to the quantum fluctuation at zero
temperature \cite{sachdev}. One dimension spin chains which may be
solved exactly are rather attractive in such studies, especially by
using the  concepts developed in quantum information theory, such as
entanglement of spins in spin chains \cite{osborn2002}, as well as
the geometric phase \cite{carollo2005} and the  fidelity
\cite{zanardi}. Other relations have already been established
between the quantum phase transitions in environment and the
decoherence of system \cite{quan2006}, entanglement \cite{yi2006},
geometric phase \cite{yuan2007, yi2007}, where the property of
system play the role of detector of the environment. These studies
have also stimulated us to consider LZ transitions in the
environment of spin chains, and such a study may provide a new
method to investigate quantum phase transitions.
Actually, the relation between LZ transitions and the quantum phase transitions
has been considered in Ref. \cite{zurek2005}, where they studied the
LZ transitions of the spin chain itself under a time-dependent evolution,
here we shall consider the LZ transition of a coupled
time-dependent two-state system and regard the chain as an environment.

In this paper, we will investigate the LZ transitions in two-state
systems surrounding by spin chains in transverse fields near their
critical points. We will focus on two kinds of spin-chain, one is
the Ising spin chain and another is XY spin chain. As we shall  show
you,  the environment's properties can also be reflected in the
systems' LZ transitions.

\section{Ising spin chain as an environment}
Taking a spin chain described by the Ising model as the environment,
we restrict ourself to consider the case where the  chain-system
coupling  only affects the spin flip of the two level system. The
Hamiltonian that governs the dynamics of the whole system (two-level
system and the chain) reads,
\begin{eqnarray}\label{Hamiltonian}
H(t){=}\frac{vt}2\sigma^z{+}\frac{\Delta}2\sigma^x
{-}{J\sum_{j=-M}^M}{(\sigma_j^z\sigma^z_{j+1}{+}\lambda\sigma^x_j{+}\frac{g}J\sigma^x\sigma_j^x)}.
\end{eqnarray}
Here $\frac{vt}2\sigma^z+\frac{\Delta}2\sigma^x$ is the standard
Landau-Zener Hamiltonian, with
$\sigma^z=|{\uparrow}\rangle\langle{\uparrow}|-|{\downarrow}\rangle\langle{\downarrow}|$,
and
$\sigma^x=|{\uparrow}\rangle\langle{\downarrow}|+|{\downarrow}\rangle\langle{\uparrow}|$,
 $|{\uparrow}\rangle$,$|{\downarrow}\rangle$ stand for the
excited and ground state of the two level system, and this part of
the Hamiltonian describes the situation where the dynamics is
restricted to two levels that are coupled by a constant tunnel
matrix element and cross at a constant velocity. $\sigma_j^x$,
$\sigma_j^y$ and $\sigma_j^z$ stand for the Pauli operators of the
spin at site $j$ in the Ising chain; $J$ and $\lambda$ describe the
strength of the spin coupling and the transverse field. The periodic
condition has been chosen to the spin chain, and we choose
$M=(N-1)/2$ for odd $N$, where $N$ is the number of the spins in the
chain. $g$ is the coupling coefficient between the two-level system
and the surrounding Ising chain.

It is convenience to continue the calculation in the interaction picture,
in order to calculate the probability of the qubit state flips due to the LZ sweep,
and we may divide the Hamiltonian into two parts as $H(t)=H_0(t)+V$,
where $H_0(t)=\frac{vt}2\sigma^z-J\sum_j(\sigma_j^z\sigma^z_{j+1}+\lambda\sigma^x_j)$
is the bit-flip free Hamiltonian and
$V=\sigma^x(\frac{\Delta}2-g\sum_j\sigma^x_j)$ describe the bit-flip interaction.

Obviously, $H_0(t)$ may be diagonalized
as long as we diagonalize the part of Ising spin chain Hamiltonian,
and this may be realized through
 Jordan-Wigner transformation $a_l=\prod_{m<l}\sigma^x_m(-\sigma_l^z+ i\sigma_l^y)/2$,
Fourier transformation,
$c_k=\frac 1{\sqrt{N}}\sum_la_l\exp(-i2\pi lk/N)$,
and Bogoliubov transformation,
$b_k=c_k\cos\frac{\theta_k}2-ic^{\dag}_{-k}\sin\frac{\theta_k}2$,
and in the diagonalize procedure, we have defined
$\cos\theta_k=\varepsilon_k/{\xi_k}$,
in which $\varepsilon_k$ and $\xi_k$ are defined as
$\varepsilon_k=2J(\lambda-\cos\frac{2\pi k}N)$,
while $\xi_k$ are defined as
$\xi_k=2J\sqrt{[\cos\frac{2\pi k}N-\lambda]^2+\sin^2\frac{2\pi k}N}$.
These transformations have transformed the spin operators
into the quasi fermion operators in the momentum space,
which may greatly simplify our following studies.
From this procedure,
we finally obtain the diagonalized expression,
$H_0(t)=\frac{vt}2\sigma^z+\sum_k\xi_kb_k^{\dag}b_k$,
following which we will get the expression of
$U_0(t)=\exp[-\frac i{\hbar}\int_{-\infty}^{t'}H_0(t')dt']$,  i.e.,
\begin{eqnarray}\label{u0t}
U_0(t)=\exp(-i\sum_k\frac{\xi_k}{\hbar}b^{\dag}_kb_kt)\cdot\exp(-\frac {ivt^2}{4\hbar}\sigma^z).
\end{eqnarray}

The Hamiltonian in the interaction picture may be obtained through
$\widetilde{H}(t)=U_0(t)^{\dag}VU_0(t)$,
and we may rewrite $V$ into the form described by $b^{\dag}_k$ and $b_k$,
which has been defined in the above diagonalize procedure,
in order to bring $\widetilde{H}(t)$ into a useful form for our following study.
Define $J^x=\sum_j\sigma^x$, which is a magnetic moment operator of the chain,
and $V=\sigma^xW$ in which $W$ is defined as
\begin{eqnarray}\label{wb}
W&=&\frac{\Delta}2-gJ^x\nonumber\\
&=&\frac{\Delta}2-g\sum_k[1-2(\cos^2\frac{\theta_k}2b_k^{\dag}b_k
+\sin^2\frac{\theta_k}2b_{-k}b_{-k}^{\dag}\nonumber\\
&+&i\frac {\sin\theta_k}2b_k^{\dag}b_{-k}^{\dag}-i\frac {\sin\theta_k}2b_{-k}b_{k})].
\end{eqnarray}

Define the product state $|{\bf{n}}\rangle=|{n_{-M}}\rangle {...} |{n_0}\rangle
|{n_1}\rangle {...} |{n_{M}}\rangle$, which stands for the state of the quasi fermion environments(Ising chain),
${\bf{n}}=\{ n_{-M}, ..., n_0, n_1, n_2,..., n_M \}$ with $n_i=0,1$,
${\bf{\Omega}}=\{ \omega_{-M}, ..., \omega_0, \omega_1, \omega_2,...,\omega_M \}$  with $\omega_k=\xi_k/{\hbar}$,
and we may change the sum $\sum_kn_k\xi_k/{\hbar}$
into the inner product form ${\bf n}\cdot{\bf \Omega}$ for simplicity.

Then, using (\ref{u0t}), (\ref{wb}) and the completeness of
$I=\sum_{\bf{n}}|{\bf{n}}\rangle\langle {\bf{n}}|$,
we may obtain $\widetilde{H}(t)$ easily,
\begin{eqnarray}
\widetilde{H}(t)&=&\sum_{\bf{m, n}}\exp[{i({\bf{m}}-{\bf{n}})\cdot{\bf{\Omega}}t}]
W_{\bf{mn}}|{\bf{m}}\rangle\langle {\bf{n}}|\nonumber\\
 &\otimes&[\exp({\frac{ivt^2}{2\hbar}})|{\uparrow}\rangle\langle {\downarrow}|
 +\exp({-\frac{ivt^2}{2\hbar}})|{\downarrow}\rangle\langle {\uparrow}|],
\end{eqnarray}
in which we have defined $W_{\bf{mn}}=\langle{\bf{m}}|W|{\bf{n}}\rangle$.

\section{Landau-Zener transition probability}
Now we can calculate the LZ transition probability based on the above preparing work.
Suppose that at time $t=-\infty$ the two-level system
is in its excited state $|{\uparrow}\rangle$
and the Ising spin chain system starts in its ground state, i.e.,
$|{\bf{0}}\rangle=\prod_{k>0}(\cos\frac{\theta_k}2+i\sin\frac{\theta_k}2c^{\dag}_{k}c^{\dag}_{-k})
|0\rangle_k|0\rangle_{-k}$ which satisfies $b_{\pm k}|{\bf{0}}\rangle=0$,
and the state of the whole system may be expressed as
$|\widetilde{\psi}(-\infty)\rangle=|{\uparrow}\rangle|{\bf{0}}\rangle$.
We now begin to calculate the survival probability
of the initial state $|{\uparrow}\rangle$ at time $t=\infty$, i.e.,
$P_{{\uparrow}{\rightarrow}{\uparrow}}(\infty){=}
|\langle {\uparrow}|\widetilde{\psi}(\infty)\rangle|^2$.
The evolution operator $\widetilde{U}(\infty,-\infty)$
with $\widetilde{U}(t_2, t_1)={\cal{T}}\exp[-\frac i {\hbar}\int_{t_1}^{t_2}d\tau\widetilde{H}(\tau)]$,
may be expressed into a time-ordered expansion,
with $t_1\leq t_2\leq ... \leq t_{2k-1}\leq t_{2k}$ in the interval $(-\infty,\infty)$,
and only the even powers of $\widetilde{H}(\tau)$
will contribute to  $P_{{\uparrow}{\rightarrow}{\uparrow}}(\infty)$.
The perturbation series for $\langle {\uparrow}|\widetilde{\psi}(\infty)\rangle$ can be expressed as,
\begin{eqnarray}\label{integ1}
\langle {\uparrow}|\widetilde{\psi}(\infty)\rangle
=\sum_{k=0}^\infty\frac 1{(i\hbar)^{2k}}\int_{-\infty}^{\infty}dt_{1}
\int_{t_1}^{\infty}dt_2...\int_{t_{2k-1}}^{\infty}dt_{2k}\nonumber\\
\sum_{{\bf{n}}^{(2k)},...,{\bf{n}}^{(1)}}
W_{{\bf{n}}^{(2k)}{\bf{n}}^{(2k-1)}}W_{{\bf{n}}^{(2k-1)}{\bf{n}}^{(2k-2)}}...
W_{{\bf{n}}^{(1)}{\bf{0}}}\nonumber\\
\cdot\exp\{\frac{iv}{2\hbar}\sum_{l=1}^{k}(t_{2l}^2-t_{2l-1}^2)
+i\sum_{l=1}^{2k}w_lt_l\}|{\bf{n}}^{(2k)}\rangle,
\end{eqnarray}
in which we have defined $w_1={\bf n}^{(1)}\cdot{\bf\Omega}$,
$w_l=({\bf{n}}^{(l)}-{\bf{n}}^{(l-1)})\cdot{\bf{\Omega}},(2\leq l\leq2k)$ for simplicity,
and ${\bf{n}}^{(l)}$ denote the state of the chain after the $l$-th interactions.

In order to work out the above integration,
it is advantageous to make the variable transformations.
Introduce a set of new variables as \cite{kayanuma1987},
$x_q=\sum_{l=1}^{2q-1}(-1)^{l+1}t_l$, $y_q=t_{2q}-t_{2q-1}$, $(1{\leq} q{\leq} k)$,
and consequently $x_l+\sum_{q=1}^{l-1}y_q=t_{2l-1}$.
Then the above perturbation series (\ref{integ1}) for
$\langle {\uparrow}|\widetilde{\psi}(\infty)\rangle$ can be changed to the form Eq. (\ref{integ}),
\begin{widetext}{
\begin{eqnarray}\label{integ}
\sum_{k=0}^\infty\frac 1{(i\hbar)^{2k}}\int_{-\infty}^{\infty}dx_1
\int_{x_1}^{\infty}dx_2
... \int_{x_{k-1}}^{\infty}dx_k
\int_{0}^{\infty}dy_1
...\int_{0}^{\infty}dy_k\sum_{{\bf{n}}^{(2k)},...,{\bf{n}}^{(1)}}
W_{{\bf{n}}^{(2k)}{\bf{n}}^{(2k-1)}}W_{{\bf{n}}^{(2k-1)}{\bf{n}}^{(2k-2)}}...
W_{\bf{n}^{(1)}{\bf{0}}}\nonumber\\
\cdot\exp\{\frac{iv}{\hbar}\sum_{l=1}^{k}y_l(x_l+\sum_{q=1}^{l-1}y_q-\frac 1 2 y_l)
+i\sum_{l=1}^k[w_{2l}y_l+(w_{2l}+w_{2l-1})(x_l+\sum_{q=1}^{l-1}y_q)]\}|{\bf{n}}^{(2k)}\rangle.
\end{eqnarray} }
\end{widetext}

An analogous analysis to references \cite{wubs2006, saito2007}
will help us to work out the above integration.
Under permutation of the $x_{l}$,  the integrand is not symmetric.
If we transform the $x_{l}$ into new variables $s_{1}=x_{1}$
and $s_{l}=x_{l}-x_{l-1}$ for $l =2,3,\ldots,k$, then $x_l=\sum_{q'=1}^ls_{q'}$.
It is easy to find that the  $\int_{-\infty}^{\infty}\mbox{d}s_{1}$-integral
will give us the $\delta$-function
$(2\pi\hbar/v)\delta(\sum_{l = 1}^{k} y_{l}+\frac{\hbar}{v}(w_{2l}+w_{2l-1})\,)$.
And since the initial state is $|{\uparrow}\rangle|{\bf{0}}\rangle$,
we can obtain $\sum_{l=1}^k(w_{2l}+w_{2l-1})={\bf n}^{(2k)}\cdot{\Omega}\ge 0$,
further more, the variables $y_{l}\geq0$,
therefore, the integral on delta-function can only exist in the subspace
$y_{1}{=}y_{2}{=}{\ldots}{=}y_{k}{=}0$ and only in the condition that the vector
${\bf n}^{(2k)}{=}{\bf 0}$.
Then we can return to the above integration with the condition of $y_l=0, (l=1,...,k)$,
and since within this subspace the integrand symmetric in the variables $x_{l}$,
we may symmetrize the $x_{l}$-integrals,
by replacing the integrals of $x_l$ to
$(1/k!)\int_{-\infty}^{\infty}dx_{1}\ldots\int_{-\infty}^{\infty}dx_{k}$.
After the $x_l$ integrals,
then the $y_{l}$-integrals can also be evaluated as well
by using the property of $\delta$-functions, $\int_0^{\infty}\delta(y)dy=1/2$.
From the above time integrals  we have noticed that
when the environment starts in the ground state $|{\uparrow}\rangle\,|{{\bf 0}}\rangle$,
only the $(2k){\mathrm{th}}$-order processes
contribute to the survival probability $P_{{\uparrow\rightarrow \uparrow}}(\infty)$,
and they should satisfy ${\bf n}^{(2)}{=}{\bf n}^{(4)}{=}\ldots{=}{\bf n}^{(2k)}{=}{\bf 0}$,
then the quasi fermion environments will end up in their initial state $|{{\bf 0}}\rangle$
in case the two-level system ends up in $|{\uparrow}\rangle$.
However, the time integrals of the equation do not prohibit the occupation  of the states
$|{\uparrow}\rangle|{{\bf n}\ne {\bf 0}}\rangle$  at intermediate times,
and they also do not restrict the intermediate environment states $|{\bf n}^{(2l-1)}\rangle$,
but the the vanishing matrix elements $W_{\bf mn}$ give the further restrictions to the integral.

According to the above analysis and calculations,
we find that $\langle {\uparrow}|\widetilde{\psi}(\infty)\rangle$ can be simplified into
$\exp(-\pi \Gamma^2/{\hbar v})|{\bf{0}}\rangle$, in which the parameter $\Gamma^2$ have been defined as
$\Gamma^2=\sum_{{\bf n}}W_{\bf{0n}}W_{\bf{n0}}=\langle {\bf 0} |W^2|{\bf{0}}\rangle$
and we may obtain the exactly LZ transition probability for
a qubit coupled to the Ising spin-chain in a transverse field,
\begin{eqnarray}\label{plz}
P_{\uparrow\rightarrow\downarrow}(\infty)=1-P_{\uparrow\rightarrow\uparrow}(\infty)=1-e^{-2\pi \Gamma^2/{\hbar v}},
\end{eqnarray}
in which the parameter $\Gamma^2$ can be work out as
\begin{eqnarray}\label{gammalz}
\Gamma^2 
&=&(\frac {\Delta}2-g\langle J^x\rangle_{\bf 0})^2+g^2(\widetilde{\Delta} J^x)^2\nonumber\\
&=&(\frac{\Delta}2-g\sum_{k>0}\cos\theta_k)^2+ g^2\sum_{k>0}\sin^2\theta_k,
\end{eqnarray}
where the term $\langle J^x\rangle_{\bf 0}=\sum_{k>0}{\cos\theta_k}$ is just
the expecting values of the magnetic moment  of the spin chain,
and $\widetilde{\Delta} J^x\equiv\sqrt{\langle (J^x)^2\rangle_{\bf 0}-\langle J^x\rangle_{\bf 0}^2}
=\sqrt{\sum_{k>0}\sin^2\theta_k}$
is the variance of the magnetic moment at the ground state $|{\bf 0}\rangle$.
In case of the coupling coefficient $g=0$,
the expression is in consist with the usual result
$P_{\uparrow\rightarrow\downarrow}(\infty)=1-\exp(-\pi\Delta^2/{2\hbar v})$,
and when $\Delta=0$, the transition are completely determined by the environment chain.

The final expressions (\ref{plz}) and (\ref{gammalz}) tell us that
the LZ transition of the central two-level system depends on both
the expecting value of the magnetic moment $J^x$ and its variance at
ground state of the spin chain. Both the expecting value and its
variance are closely related to the strength of the transverse
field, which will affect the LZ transition equation through
$\theta_k$, thus we may conjecture  that the message of quantum
phase transition happens at the certain strength of the transverse
field may also be reflected on the LZ transition probability of the
two-level system.
\begin{figure}
\includegraphics*[width=0.45\columnwidth,
height=0.32\columnwidth]{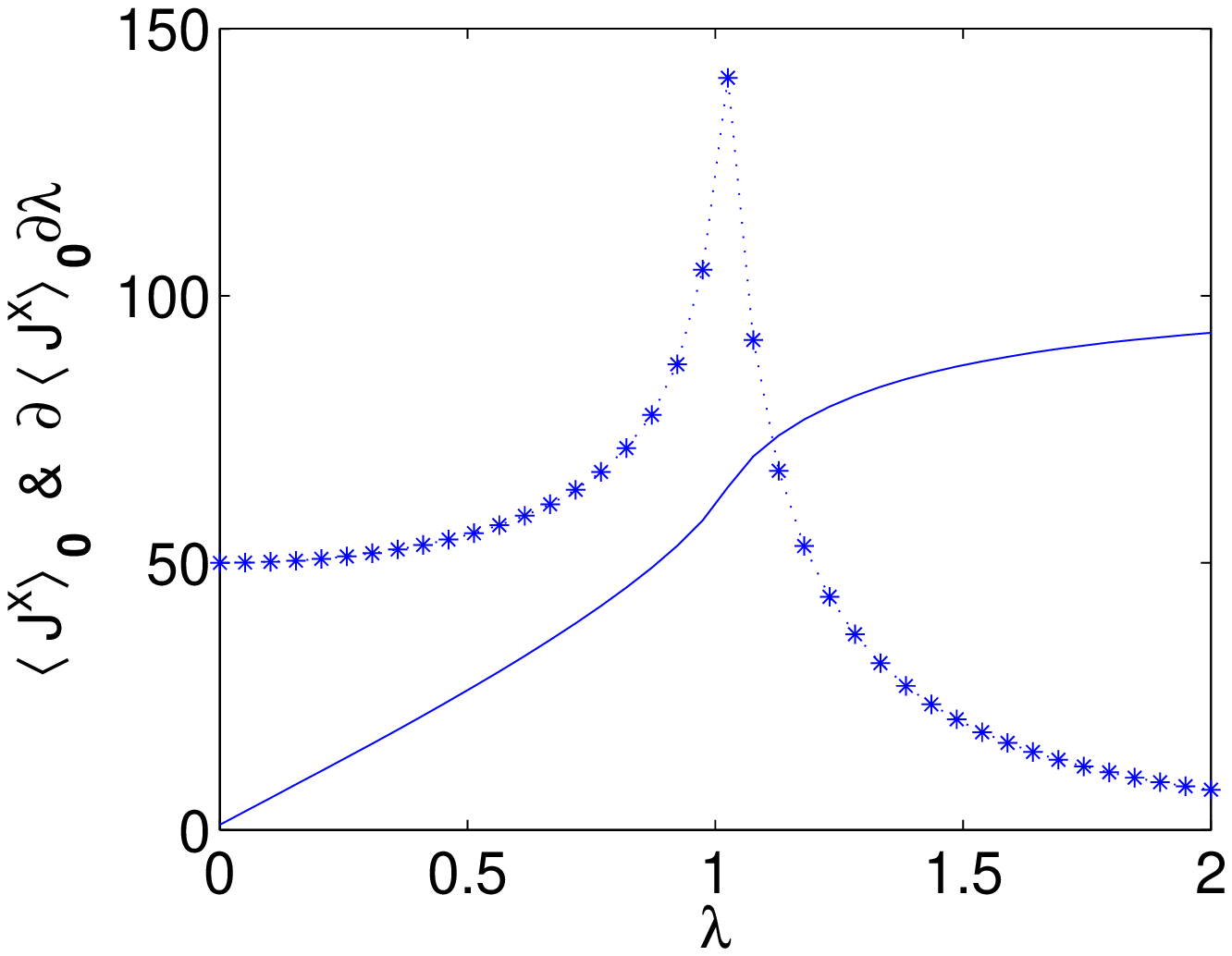}
\includegraphics*[width=0.45\columnwidth,
height=0.32\columnwidth]{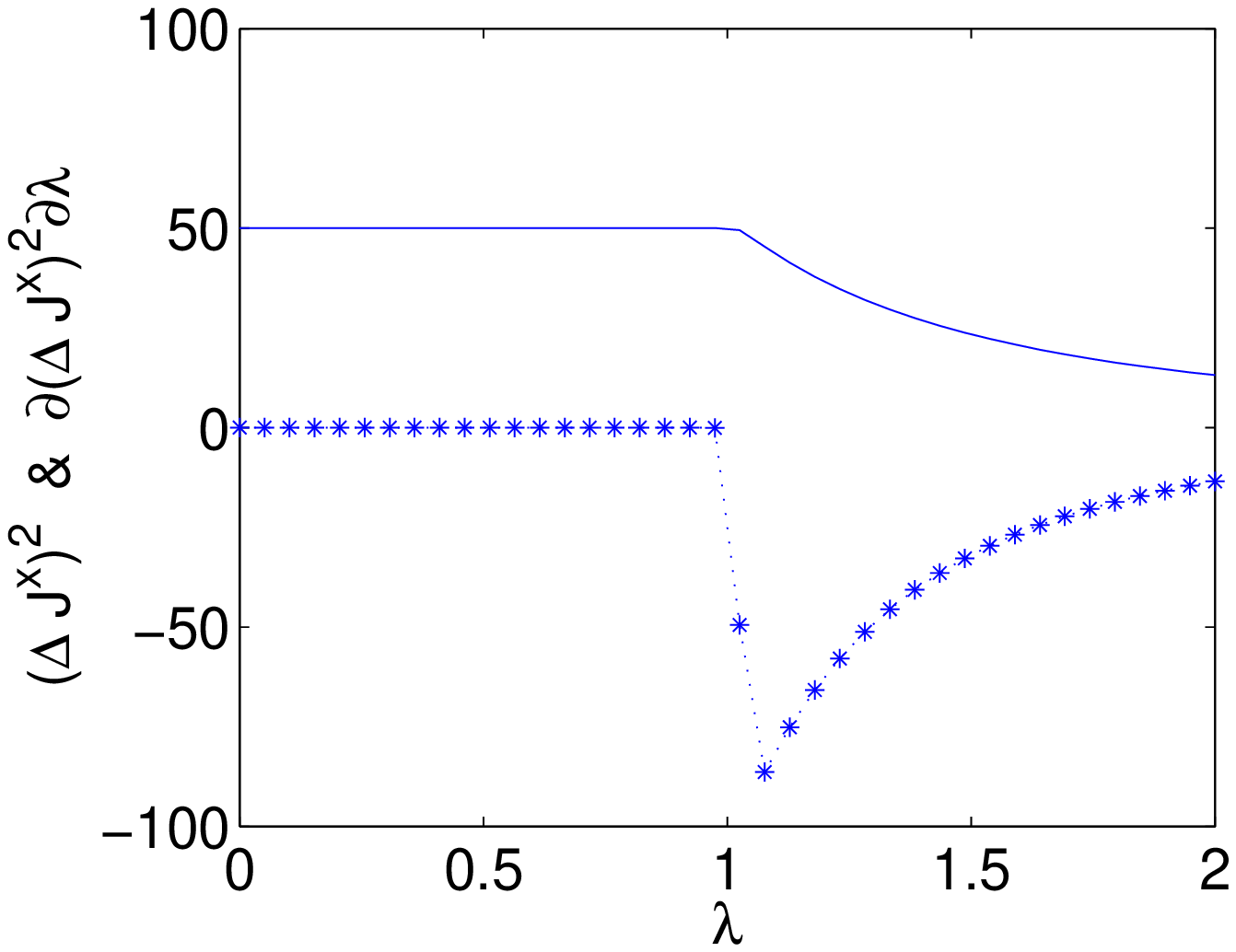}
\caption{(color online)
$\langle J^x\rangle_{\bf 0}$ and $(\widetilde{\Delta}J^x)^2$(solid line)
as well as their derivatives by $\lambda$ (dot-star line) as the functions of the parameter $\lambda$.
We set $N=200$ 
in the numerical calculation.}
\label{fig1}
\end{figure}

Fig. {\ref{fig1}} shows  $\langle J^x\rangle_{\bf 0}$ and
$(\widetilde{\Delta}J^x)^2$  as functions of the transverse field
strength $\lambda$, as well as their derivatives with respect to
$\lambda$. When the strength of the transverse field is in the
region $0{\leq}\lambda{\leq}1$, only the  expectation value  of
$J^x$ changes, while the variance part of $\Gamma^2$ does not. This
feature tells us that, the change of  LZ transitions mainly caused
by the expectation value  of magnetic moment in this case; however,
when $\lambda{\geq}1$, the variance become decreased  with the
transverse field, both of them will affect the LZ transition and the
expectation values of magnetic moment may become the mainly causation.
 The expectation value changes
sharply when $\lambda \rightarrow 1$, indicating that their
derivatives may reveal the singularity near the critical points
perfectly.
\begin{figure}
\includegraphics*[width=0.48\columnwidth,
height=0.36\columnwidth]{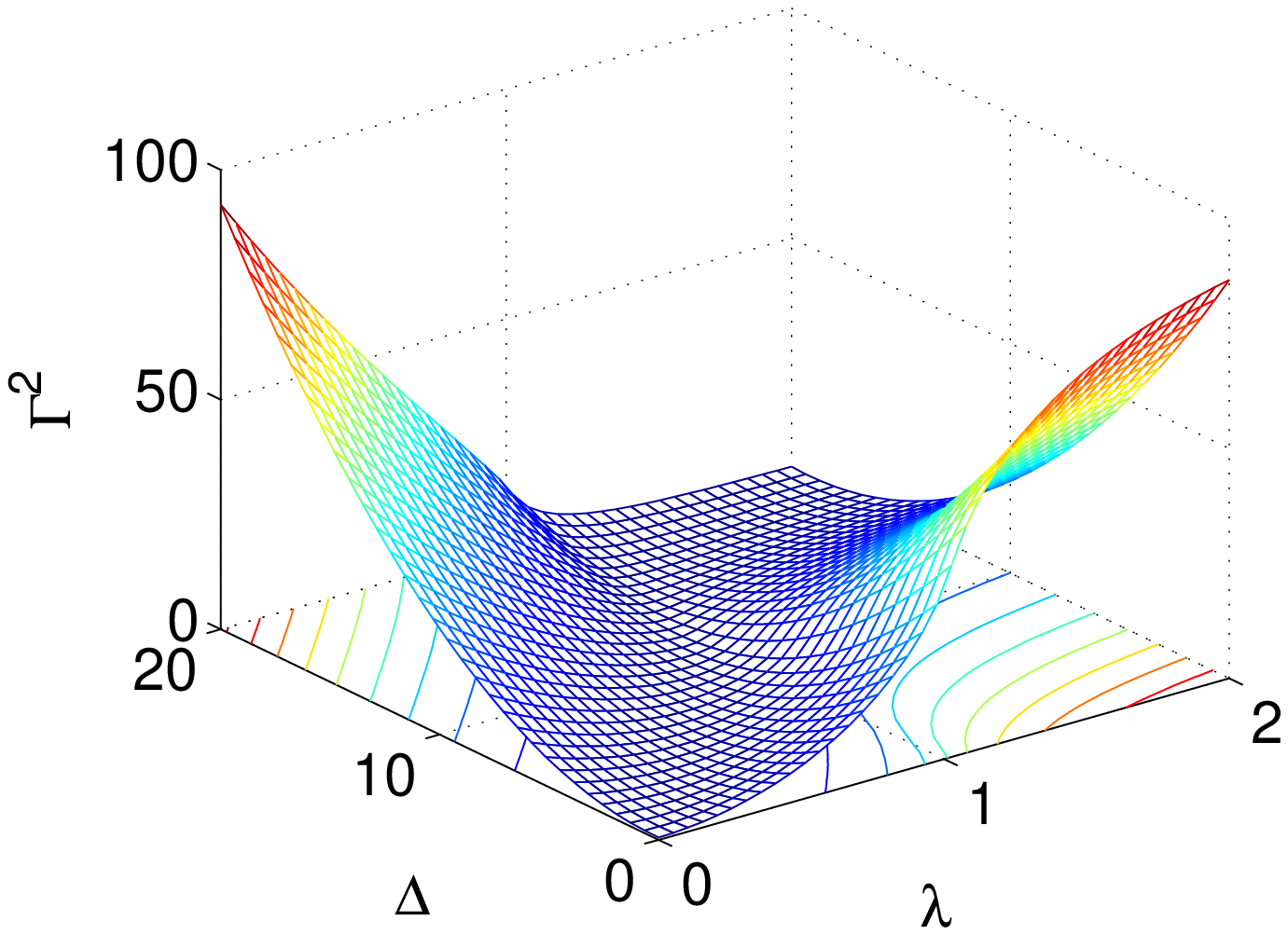}
\includegraphics*[width=0.48\columnwidth,
height=0.36\columnwidth]{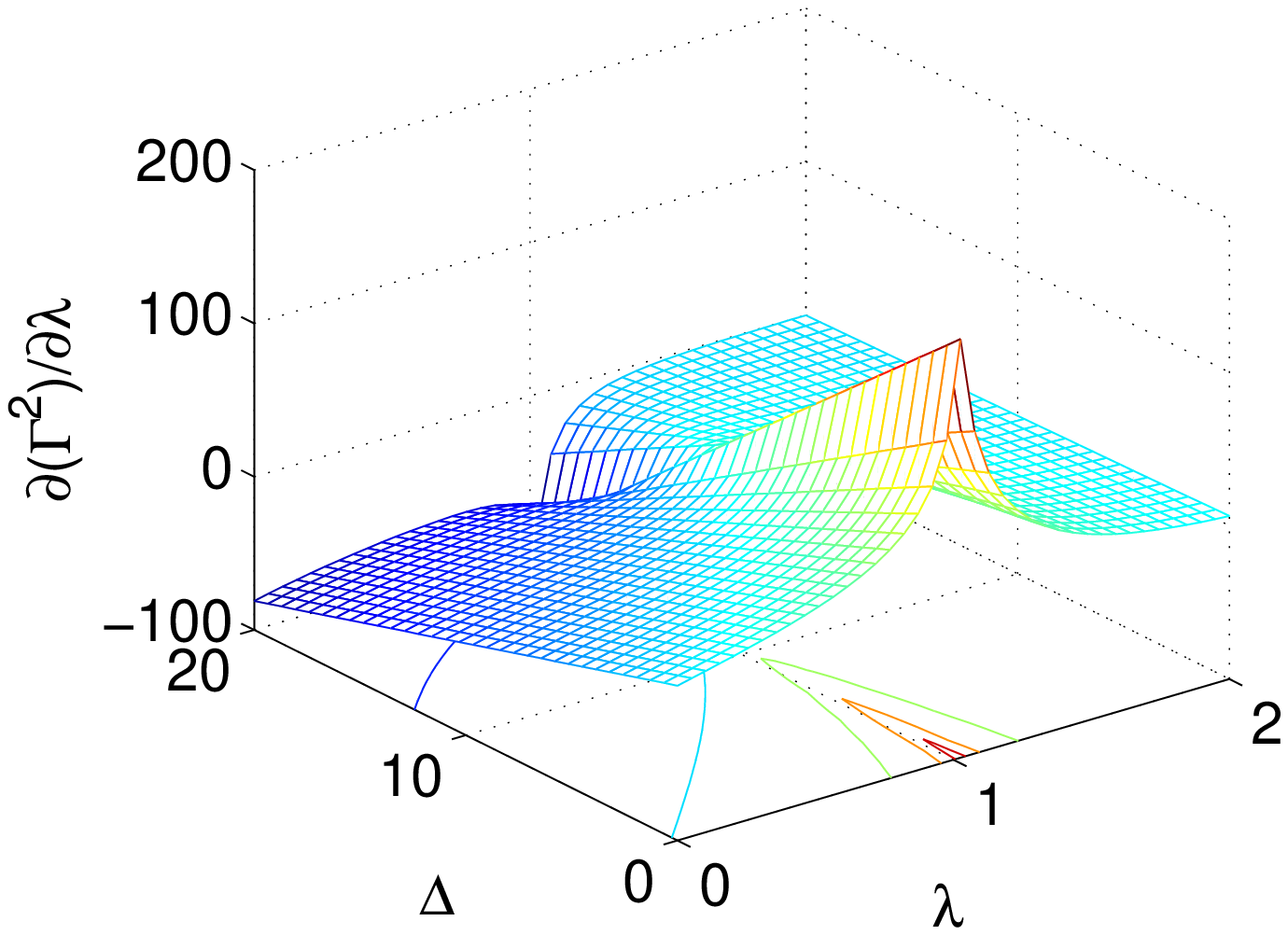} \caption{(color online) In
units of $J^2$, $\Gamma^2$ and its derivative $\partial
\Gamma^2/\partial\lambda$ as the functions of $\lambda$ and $\Delta$
. We set $N=200$, $g=0.1J$, $\Delta_{\max}=20J$, in the numerical
calculation. }\label{fig2}
\end{figure}
The above property  becomes more clear  when $N$ increases, this
result comes from the analysis given  in \cite{carollo2005,
quan2006, yuan2007}. In the following we shall present the results
with a specific number $N=200$.

\begin{figure}
\includegraphics*[width=0.48\columnwidth,
height=0.36\columnwidth]{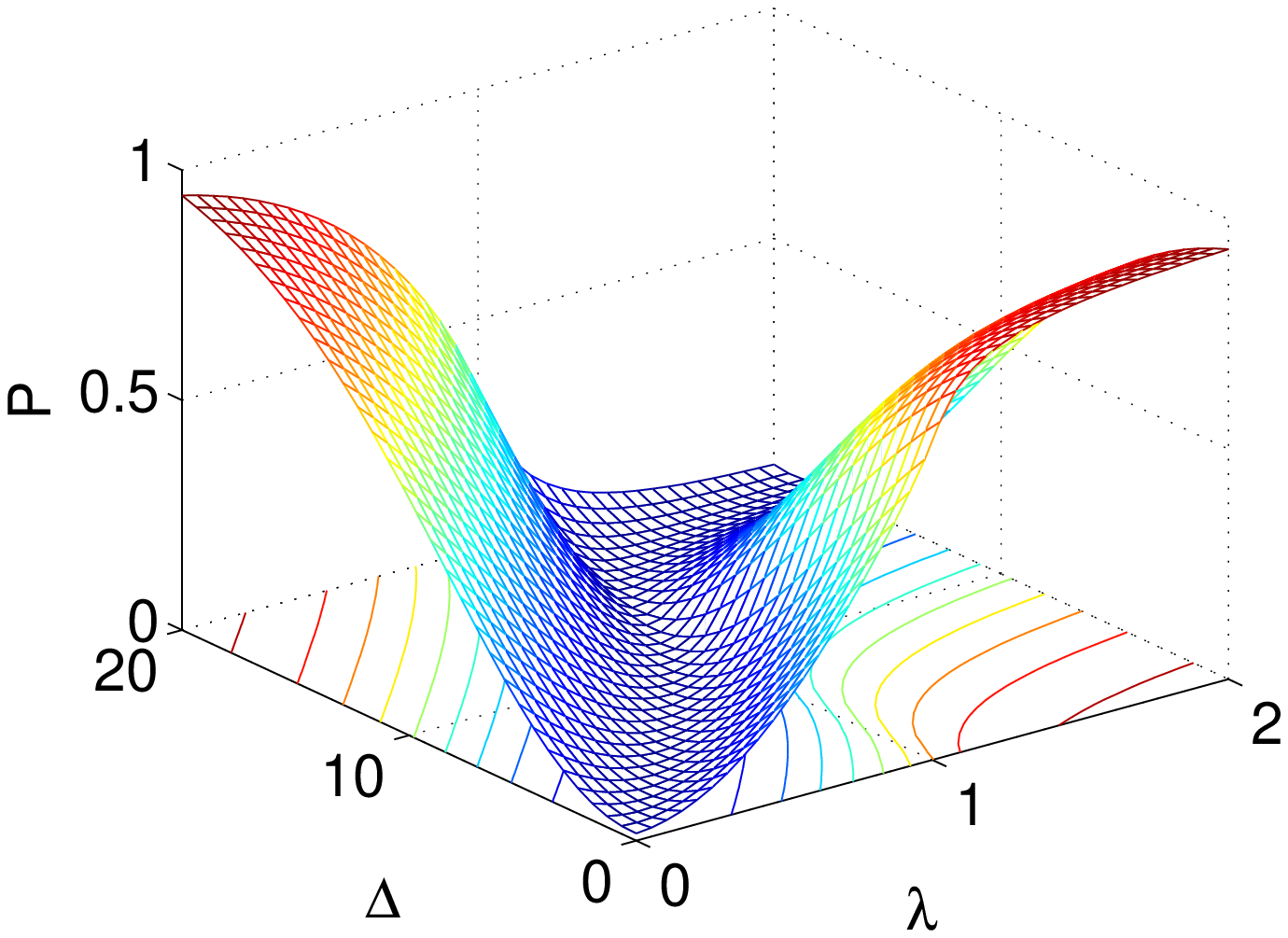}
\includegraphics*[width=0.48\columnwidth,
height=0.36\columnwidth]{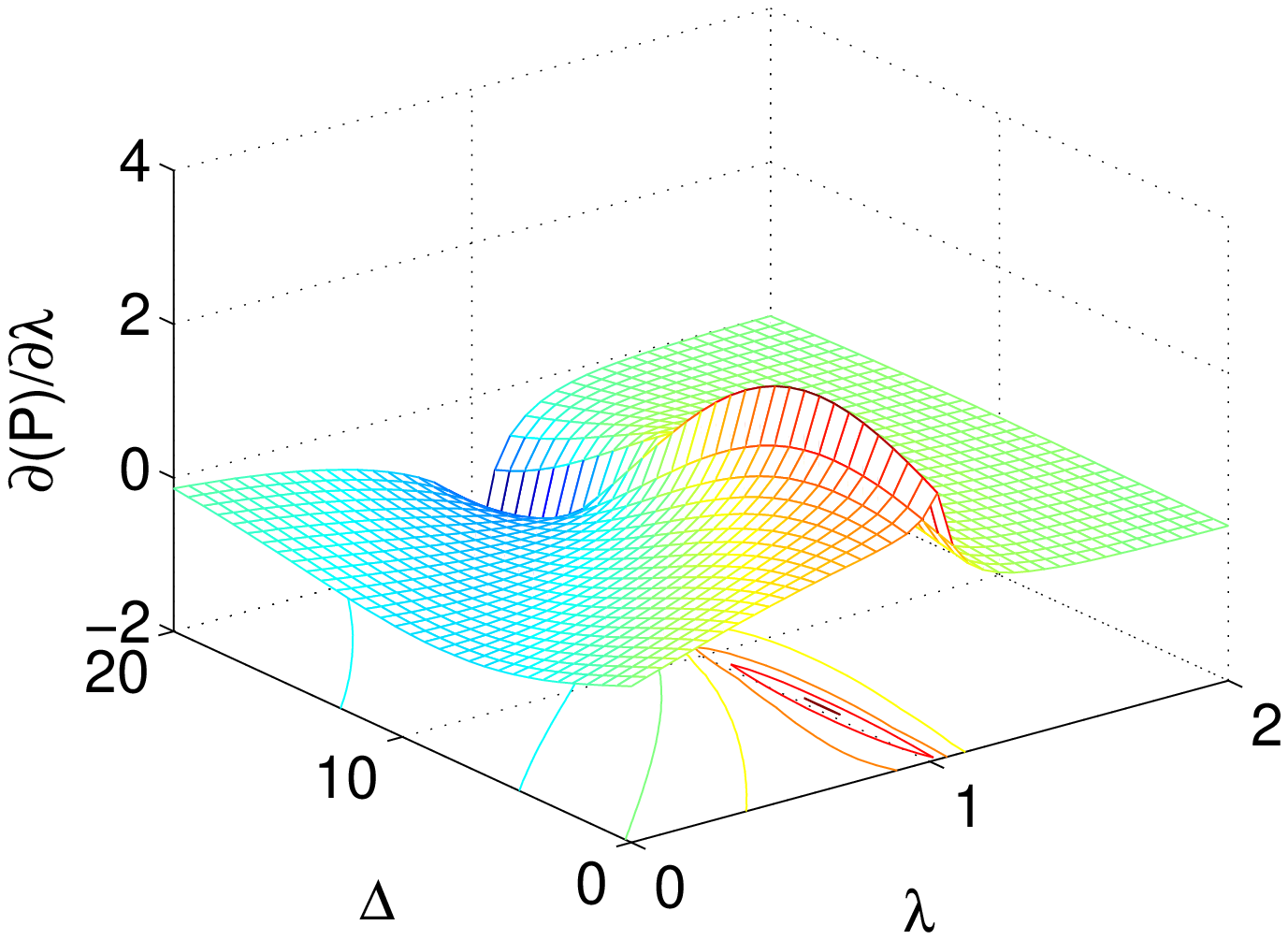}
\caption{(color online)
LZ transition probability $P$ and its derivative
$\partial P/\partial\lambda$
as the function of $\lambda$ and $\Delta$.
We set $N=200$, $g=0.1J$, $\Delta_{\max}=20J$,
$v=50{ J^2/\hbar}$ in the numerical calculation.
}\label{fig3}
\end{figure}

Since the magnetic moment of the spin chain changes fast at the critical points of the chain,
then we may expected that the LZ transition can also reflect such property of the spin chain
by the expression of (\ref{plz}) and  (\ref{gammalz}).
Fig. \ref{fig2} shows the function of $\Gamma^2$ as well as its derivative relation with $\lambda$ and $\Delta$,
we set the number of the chain $N=200$, $g=0.1J$ in the numerical illustration.
It is evident that when the strength of the transverse field $\lambda \rightarrow 1$,
$\Gamma^2$ also changes sharply.
As mentioned above, in order to reveal this effect of quantum phase transitions more clearly,
we may pay more attention to the derivative of $\Gamma^2$ by $\lambda$,
which perfectly shows the quantum critical phenomenon.
Just as we have expected, this property has been inherited very well by the LZ transition probability
$P_{\uparrow\rightarrow\downarrow}(\infty)$ and its derivative,
as Fig. \ref{fig3} shows, where we have set $v=50{ J^2/\hbar}$ in the calculation.
It is in evidence that the critical point is
reflected perfectly well in the derivative of LZ transition,
which is consist with the above analysis.

Different values of $\Delta$ also affect the property of LZ
transition probability. From Eq. (\ref{gammalz}) and Fig.
\ref{fig3}, we may find that when $\Delta>0$, the transition
probability declines first, and then reveals again, however, the
sharply changed location is not influenced, which can also give us a
good reflection of quantum phase transitions. When $\Delta$ is large
enough, the transition probability will not reveal again in the case
of $\lambda>1$.

\section{The XY spin chain as the environment}
Now we consider the case where the XY spin chain acts as the
environment. Straightforward calculation shows that to get the
result in this situation,  we need to replace
$-J\sum_{j=-M}^M(\frac{1+\gamma}2\sigma^z_j\sigma^z_{j+1}
+\frac{1-\gamma}2\sigma^y_j\sigma_{j+1}^y+\lambda\sigma^x_j)$ with
$-J\sum_{j=-M}^M(\sigma^z_j\sigma^z_{j+1}+\lambda\sigma^x_j)$ in Eq. (\ref{Hamiltonian}).
Here $\gamma$ measures the anisotropy in XY spin-chain. The XY
Hamiltonian will turn into the transverse Ising chain for
$\gamma=1$, and the XX chain in transverse field for $\gamma=0$.

\begin{figure}
\includegraphics*[width=0.48\columnwidth,
height=0.36\columnwidth]{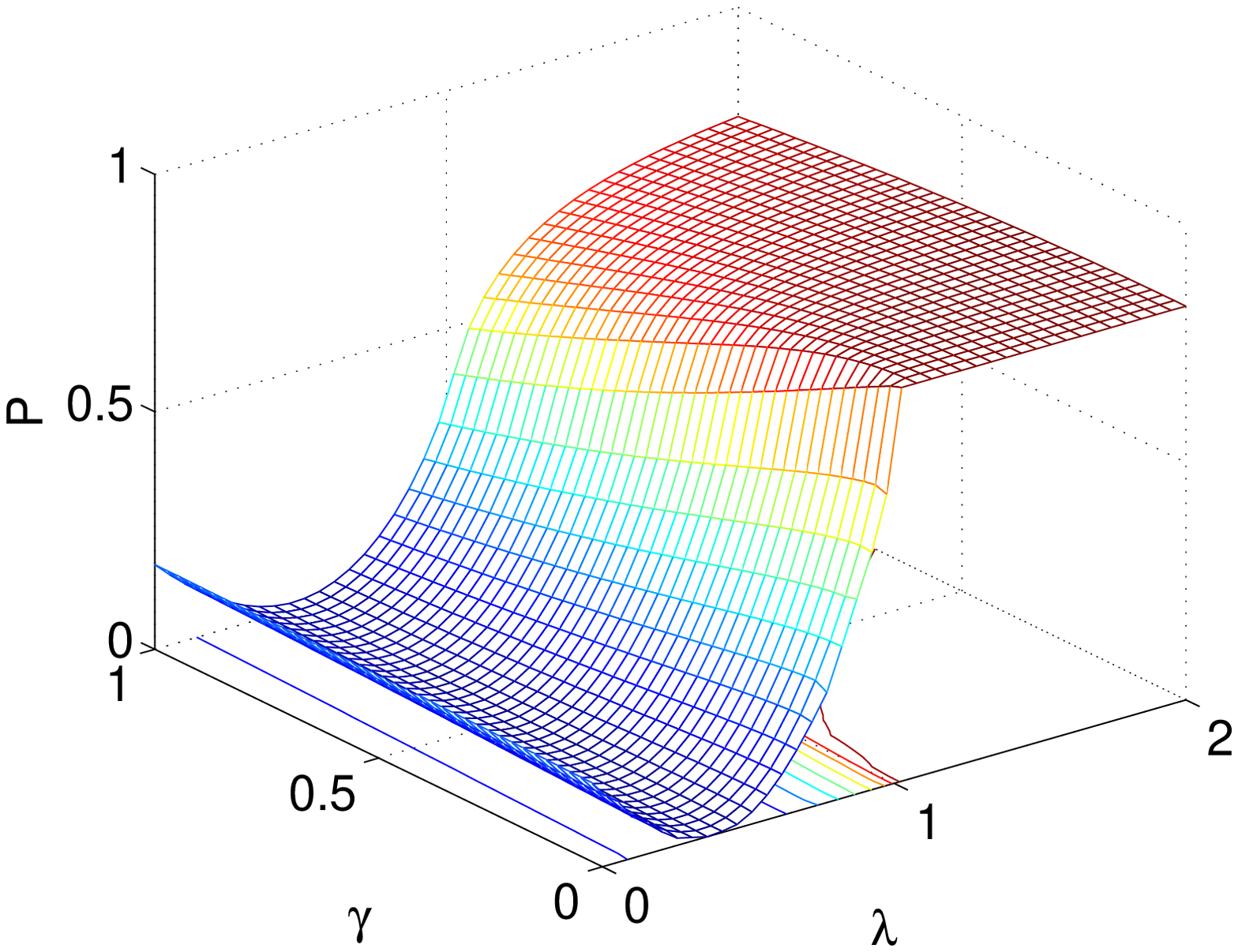}
\includegraphics*[width=0.48\columnwidth,
height=0.36\columnwidth]{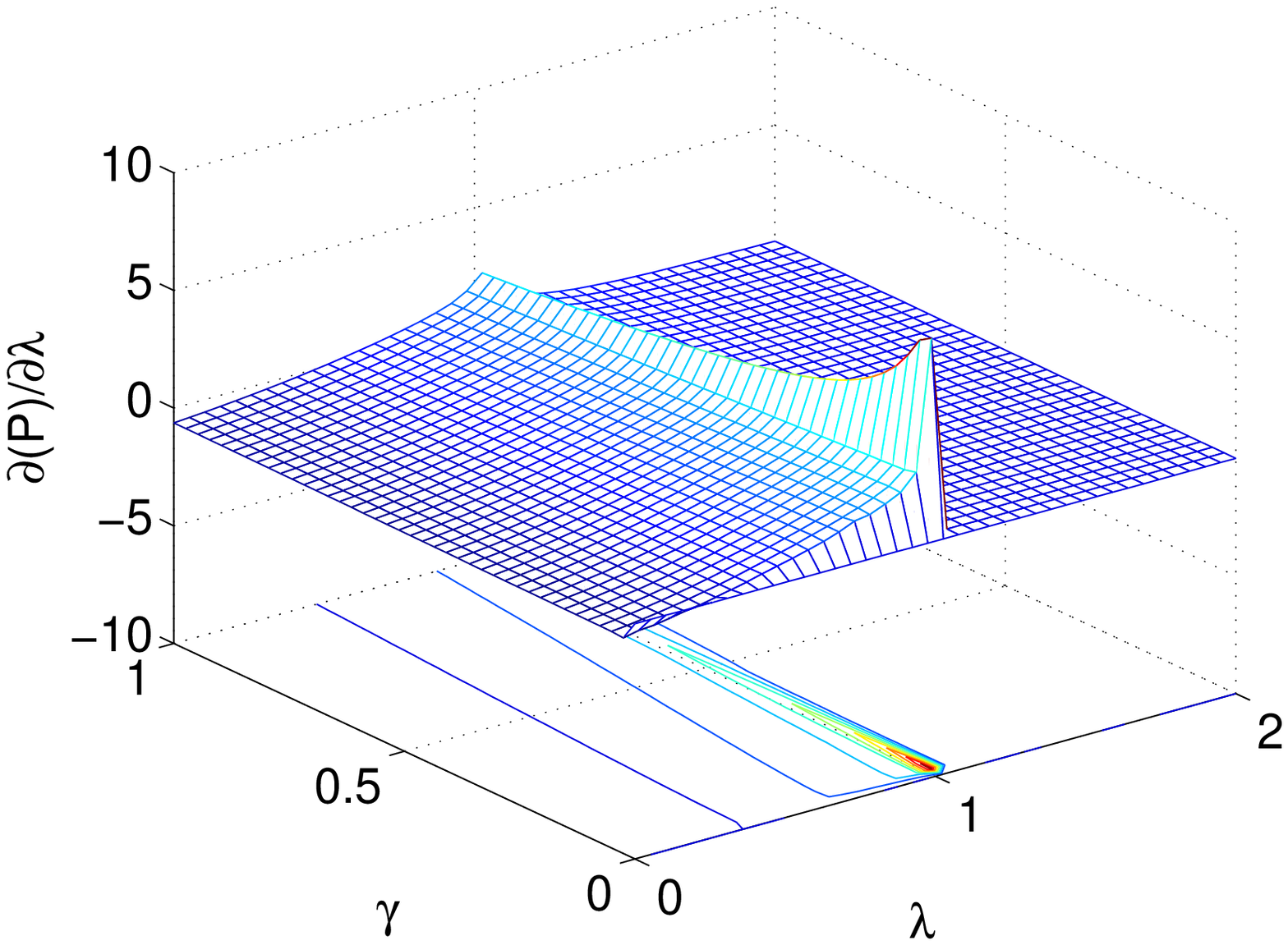}
\includegraphics*[width=0.70\columnwidth,
height=0.32\columnwidth]{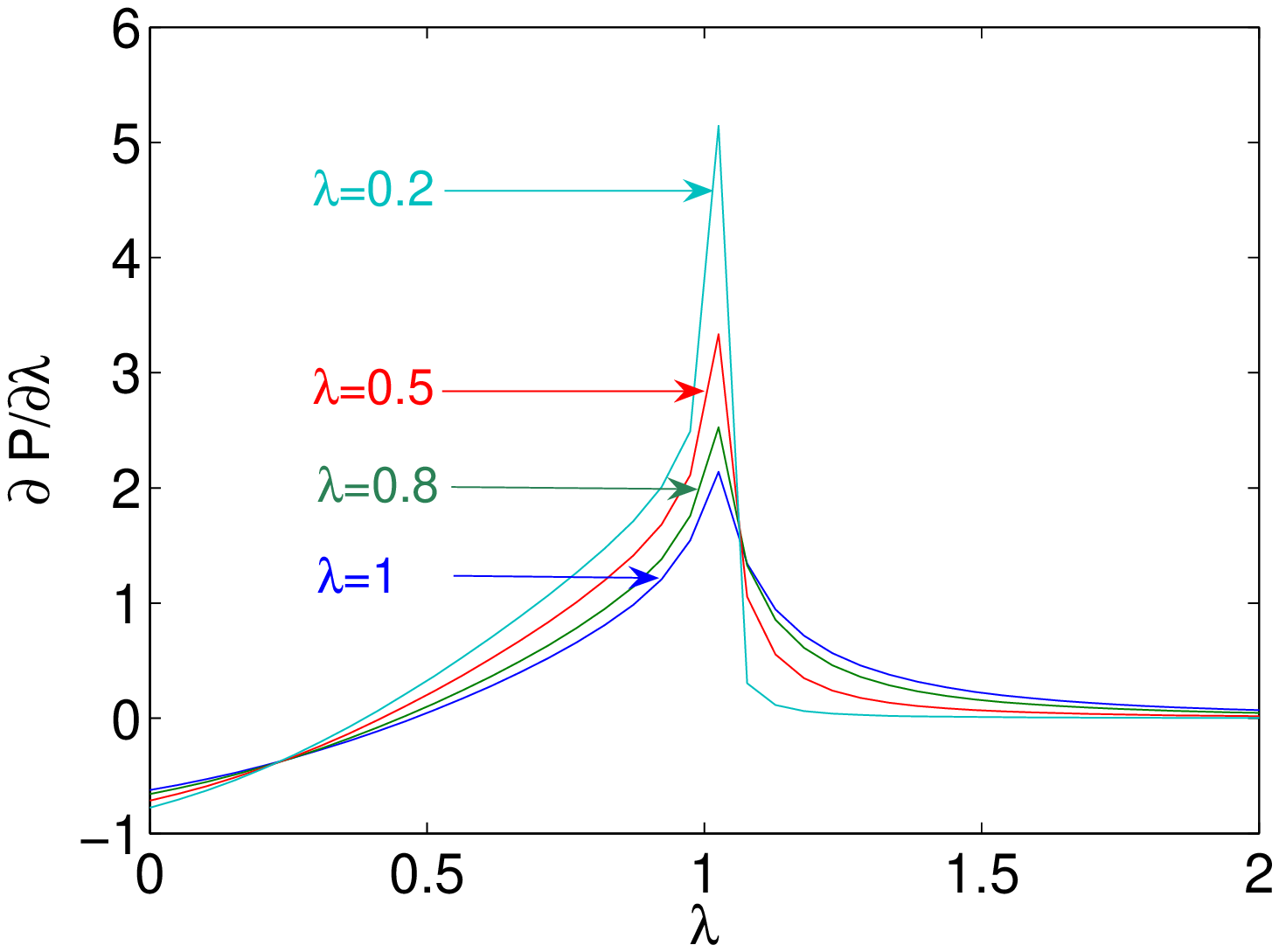}
\caption{(color online)
In the environment of XY spin chain, LZ transition probability $P$ and its derivative
$\partial P/\partial\lambda$
as the function of $\lambda$ and $\gamma$.
We set $N=200$, $g=0.1J$, $\Delta=5J$,
$v=50{ J^2/\hbar}$, in the numerical calculation.
The panel shows $\partial P/\partial\lambda$ changes with $\lambda$
in case of $\gamma=0.2,0.5,0.8,1.0$.
}\label{fig4}
\end{figure}

By the same procedure, we can obtain the LZ transition probability
in this case, the equations are nothing but  changing the definition
of $\theta_k$ and $\xi_k$ in the above discussions, i.e.,
$\cos\theta_k=\varepsilon_k/{\xi_k}$, in which
$\varepsilon_k=2J(\lambda-\cos\frac{2\pi k}N)$ and $\xi_k$ are now
defined as $\xi_k=2J\sqrt{[\cos\frac{2\pi
k}N-\lambda]^2+\gamma^2\sin^2\frac{2\pi k}N}$.

Following the same calculation, we then obtain the same expression
as (\ref{plz}) and (\ref{gammalz}). Fig. \ref{fig4} shows the
relation between LZ transition probability $P$ and the anisotropy
parameter, as well as its derivative $\partial P/{\partial\lambda}$.
We set $N=200$, $g=0.1J$, $\Delta=5J$, $v=50{ J^2/\hbar}$, in the
numerical calculation. It shows that when $\gamma$ changes from $1$
to $0$, the critical behavior is  reflected quite well in the LZ
transition, and the critical line $\lambda=1$ is clearly reflected
in the derivative of the LZ transition probability.
We may deduce that the distinctive feature of the XX chain environment
is a sharp discontinuity in the derivative $\partial P/{\partial\lambda}$
at $\lambda=1$, this is reflected more clearly from the panel in Fig. \ref{fig4}.
This result confirm our
above prediction that the LZ transition can reflect the critical
points of the environment.

\section{Conclusion}
In summary, we have studied the Landau-Zener transitions in
two-state systems coupling to  the  Ising spin chain and XY spin
chain in transverse fields. We have calculated the exact expressions
of the LZ transition probabilities of the two-state systems and
analyzed the relation between their properties and the occurrence of
the quantum phase transitions in the chain. The results show that
the  LZ transition are determined by the spin chains' magnetic
moments and  their variance. As the magnetic moments of the chains
contain the information  of quantum phase transitions, the LZ
transitions may  act as the witnesses of quantum phase transitions
in the chains. Our results suggest a rather intriguing relationship
between LZ transitions and the environments' properties, and
therefore the results may provide a new way to study the phenomenon
of quantum phase transition as well as Landau-Zener transition.

\acknowledgments
This work was supported by NCET of M.O.E, and NSF
under grant  No. 60578014.

\end{document}